\documentclass[journal]{IEEEtran}

\usepackage{cite}
\usepackage{graphicx}
\graphicspath{{fig/}}
\usepackage{amsmath}
\usepackage{algorithmic}
\usepackage{array}
\usepackage[caption=false,font=footnotesize]{subfig}
\usepackage{url}

\usepackage{amssymb}
\usepackage{amsthm}

\usepackage[draft,inline,nomargin]{fixme}
\fxsetup{theme=color,mode=multiuser}
\FXRegisterAuthor{cwh}{acwh}{\color{red}cwh}
\FXRegisterAuthor{aa}{aaa}{Other}

\begin{document}

\title{Mobile Traffic Offloading with Forecasting using Deep Reinforcement Learning}

\author{\IEEEauthorblockN{Chih-Wei Huang and Po-Chen Chen}\\
\IEEEauthorblockA{Department of Communication Engineering,
National Central University, Taoyuan, Taiwan\\
Email: cwhuang@ce.ncu.edu.tw, 106523046@cc.ncu.edu.tw}
}


\maketitle

\begin{abstract}
With the explosive growth in demand for mobile traffic, one of the promising solutions is to offload cellular traffic to small base stations for better system efficiency. Due to increasing system complexity, network operators are facing severe challenges and looking for machine learning-based solutions. In this work, we propose an energy-aware mobile traffic offloading scheme in the heterogeneous network jointly apply deep Q network (DQN) decision making and advanced traffic demand forecasting. The base station control model is trained and verified on an open dataset from a major telecom operator. The performance evaluation shows that DQN-based methods outperform others at all levels of mobile traffic demand. Also, the advantage of accurate traffic prediction is more significant under higher traffic demand.
\end{abstract}

\begin{IEEEkeywords}
Deep learning, HetNet, mobile traffic offloading, mobile traffic forecasting, multitask learning, big data.
\end{IEEEkeywords}

\section{Introduction}
%


The demand for mobile data traffic has been rapidly growing over recent years and continuing at an even faster pace. One of the promising solutions to support the massive amount of mobile data is offloading the traffic from macrocells to technologies such as small cells and WiFi\cite{aijaz2013survey}. Therefore, the heterogeneous network (HetNet) formed by a large number of cloud-assisted small cells is one of the 5G dominant themes to enhance overall network capacity. Due to increasing complexity, network operators are facing severe challenges on inter-cell resource management\cite{Zhang2015}.

Utilizing the collected transmission records, cellular network operators can deploy a more effective self-organizing network (SON) engine for a wide range of management tasks\cite{Imran2014}. To deal with changing operational environments, the 5G SON covers both short and long-term dynamics adapting to data usage patterns over hours and days. Instead of being reactive, SON requires to be proactive with the ability to predict the performance issues and take preemptive actions before damage occurs. Though solutions such as dynamically switching off base stations (BSs) are studied\cite{Han2016}, more proficient strategies remain to be investigated.

At the same time, machine learning-based approaches have been proven to interact with complex environments successfully and emerging for the fifth-generation (5G) wireless networks on prediction and management tasks\cite{Jiang2017}. However, the application and end-device dependent behaviors result in the time-varying nature of mobile data traffic, which is hard to be modeled\cite{Shafiq2011}. The deep learning techniques have been applied to discover the latent features representing the dynamics of Internet traffic flow\cite{Oliveira2016, Cortez2012, Huang2017}.
For mobile network offloading, both optimization theory and reinforcement learning (RL) based solutions have been actively proposed\cite{Saker2012,chen2015energy,Simsek2015,Sun2015,Zhang2016,Wei2018}. The ON/OFF status of small BSs can be selected considering energy efficiency. Various learning-based schemes also determine more frequent actions such as power allocation, user association, and scheduling. Nevertheless, there is seldom research integrating mobile traffic offloading with traffic flow prediction and implemented on a realistic dataset.

In this work, we propose an energy-aware mobile traffic offloading scheme in HetNet jointly apply deep Q network (DQN) based decision and traffic demand forecasting. The DQN mitigates the challenge of continuous state space and discrete action space in traditional reinforcement learning by neural networks\cite{mnih2015human}. The proposed network management scheme offloads the transmitting data from macro to small BSs, considering traffic and performance provisioning for improved energy efficiency. Focusing on the machine learning-based approach for SON, we contribute:
\begin{itemize}
    \item A novel energy-aware architecture for mobile traffic offloading using DQN. Instead of raw traffic values, provisioning statistics required to reach offloading decisions are learned for better performance and training efficiency.
    \item Integration with advanced traffic forecasting\cite{Huang2017}. With the proven combination of convolutional neural network (CNN), recurrent neural network (RNN), and multitask learning (MTL), the results are further enhanced.
    \item Training on a big open dataset\cite{Barlacchi2015} from a major telecom operator. Real Internet activities and macro BS locations are processed to support the HetNet offloading environment.
\end{itemize}
The simulations show DQN-based methods outperform referenced Q-learning and others at all levels of mobile traffic demand due to the ability to handle complex environments. Also, traffic forecasting provides clear advantages and plays a more critical role when the macrocells are heavily loaded.
Overall, the proposed pre-possessing and DQN model featuring traffic statistic prediction can significantly enhance the power of DRL for HetNet traffic offloading.

The rest of this paper is organized as follows.
Section~\ref{sec:related} describes the related research work.
Section~\ref{sec:system} introduces the system model.
Section~\ref{sec:offloading} addresses the details of the DQN based traffic offloading scheme.
Section~\ref{sec:data} explains the data processing with public datasets.
The simulation results are discussed in Section~\ref{sec:results}.
Finally, conclusions are drawn in Section~\ref{sec:conclusion}.

\section{Related Work}\label{sec:related}

\subsection{Deep Reinforcement Learning}


RL is a computational approach allowing an agent to find a suitable action for the environment so that the reward can be maximized\cite{sutton1998reinforcement}. The agent, namely the learner and decision-maker, observes the state of the environment and acts at each decision epoch. The environment feedbacks a reward to the agent and enters the next epoch. Therefore, the agent learns the properties of the environment by interacting with it. The goal of an RL agent is to maximize its expected total reward. Q-learning is a model-free RL technique that uses unsupervised training to learn an unknown environment\cite{watkins1992q}. Deep reinforcement learning (DRL) significantly enhances the capability of RL with deep neural networks (DNN). As for DQN, the Q-functions are approximated by DNNs to deal with a large number of states\cite{mnih2015human}. In this work, we explore the potential of DQN to enhance network management tasks with discrete action space and find that it is capable of providing satisfactory results.

\subsection{Traffic Offloading in HetNet}

To solve the energy-efficiency oriented traffic offloading problem in HetNets, researchers have proposed various methods to change the operating states (ON and OFF) dynamically of small BSs\cite{Han2016}.
Specific topics range from energy/power consumption\cite{marsan2012multiple} and joint optimization of energy consumption and data throughput\cite{soh2013energy}.
Liu at el.\cite{liu2016small} presented a stochastic geometry-based model to solve the trade-off between data throughput and energy consumption. Also, user activity forecasting was applied to enhance mobile data offloading. The procedure in\cite{siris2013performance} offloaded mobile traffic to WiFi hotpots exploiting mobility prediction and prefetching considering delay and energy consumption.
Zhang et al.\cite{Zhang2016} jointly optimized user association, small BS ON/OFF states, and transmission power in green HetNet. A closed-form solution was derived for the single small BS case, while a two-stage energy-aware scheme solved the multiple BS case.

Recently, RL based mobile traffic management schemes attract growing attention.
Authors in\cite{UlIslam2012} used classic Q-learning to deal with the coverage and capacity optimization (CCO) problem in SON.
Bennis et al.\cite{Bennis2013} modeled the sub-carrier selection of small and macro cells as a non-cooperative game aiming to mitigate the interference caused to the macrocell network while maximizing spectral efficiency. The RL-based procedure converges to an epsilon Nash equilibrium when all small cells share the same interests.
Simsek et al.\cite{Simsek2015} proposed a mobility management approach in HetNet, which jointly learned long-term traffic load balancing and optimal cell range expansion. The multi-armed bandit learning was adopted in the algorithm design.
Authors in\cite{Sun2015} dealt with the traffic offloading issue in two-tier multi-mode small cell networks over unlicensed bands. The proposed hierarchical Q-learning framework can discover a mixed strategy equilibrium of the corresponding Stackelberg game.
Chen et al.\cite{chen2015energy} formulated the energy-aware traffic offloading problem as a discrete-time Markov decision process (DTMDP), whose statistics depend on the traffic offloading strategy. The authors provided a model-free Q-learning technique with state aggression to solve the optimal traffic offloading strategy.
In\cite{Wei2018}, the authors investigated the user scheduling and resource allocation policy in HetNet with a hybrid energy supply. The policy gradient-based actor-critic algorithm was adopted to deal with continuous-valued states and actions. A probabilistic model and linear function approximation were used for actor and critic parts, respectively.
Authors in\cite{Chinchali2018} developed a data-driven network simulator to study IoT traffic scheduling on top of conventional applications. A DRL-based approach was proposed to provide control decisions in homogeneous networks.
Therefore, HetNet traffic offloading using a DRL-based approach with realistic data is less discussed and worth further study.

\subsection{Traffic Forecasting}

Machine learning and time series analysis has been an emerging tool to model and predict network traffic.
In\cite{Xu2016}, the authors analyzed traffic patterns of 9000 cellular towers in a metropolitan area. The study decomposed mobile traffic into regular components and unpredictable random components.
After analyzing thousands of base stations and billions of records, Zhou et al.\cite{Zhou2012} classified mobile traffic into three application types and concluded that knowledge of adjacent cells enhanced the traffic prediction performance.
Li et al.\cite{Li2017} proposed an application-level traffic prediction framework for cellular networks. The model used an $\alpha$-stable model to capture the spatial and temporal characteristics of applications. Then dictionary learning was implemented to refine coarse prediction results. Prediction accuracy was improved by applying customized fitting models to different traffic types.

Deep learning-based approaches have also been studied recently to discover possible features from vehicular and Internet traffic flows.
In\cite{Lv2015}, a stacked autoencoder model was applied to learn features from vehicular traffic based on data collected from freeways. The prediction scheme with deep learning outperformed other methods.
Huang et al.\cite{Huang2014} proposed a useful MTL architecture which consists of a deep belief network for feature extraction and a multitask regression stage for supervised prediction.
Oliveira et al.\cite{Oliveira2016} studied algorithms based on artificial neural networks for general Internet traffic forecasting. The recurrent RNN performed better than stacked autoencoder and is suitable for time-series network traffic prediction.
Traffic forecasting on a public dataset with the latest deep learning model was proposed in our previous work\cite{Huang2017}.
The CNN-RNN model takes full advantage of machine learning toward intelligent network management in 5G systems.

\section{System Model}\label{sec:system}

This section describes the environment as well as the system architecture of mobile traffic offloading in HetNet. Figure 1 presents the environment where a cloud controller connects to multiple macro and low-power small BSs forming macro and small cells. When a small BS is activated, it cooperates with the macro BS to mitigates the data traffic loads. A macro BS consumes more energy at higher loading rates, so offloading traffic to small BSs may reduce energy consumption. By contrast, if a macro BS is at low loading rates, turning on small BSs can degrade energy efficiency. An adaptive offloading strategy is needed to develop an energy-efficient HetNet system under dynamic mobile traffic. The proposed architecture predicts the mobile data traffic in advance and decides the number of active small BSs considering energy efficiency.

\begin{figure}
	\centering
	\includegraphics[width=0.4\textwidth]{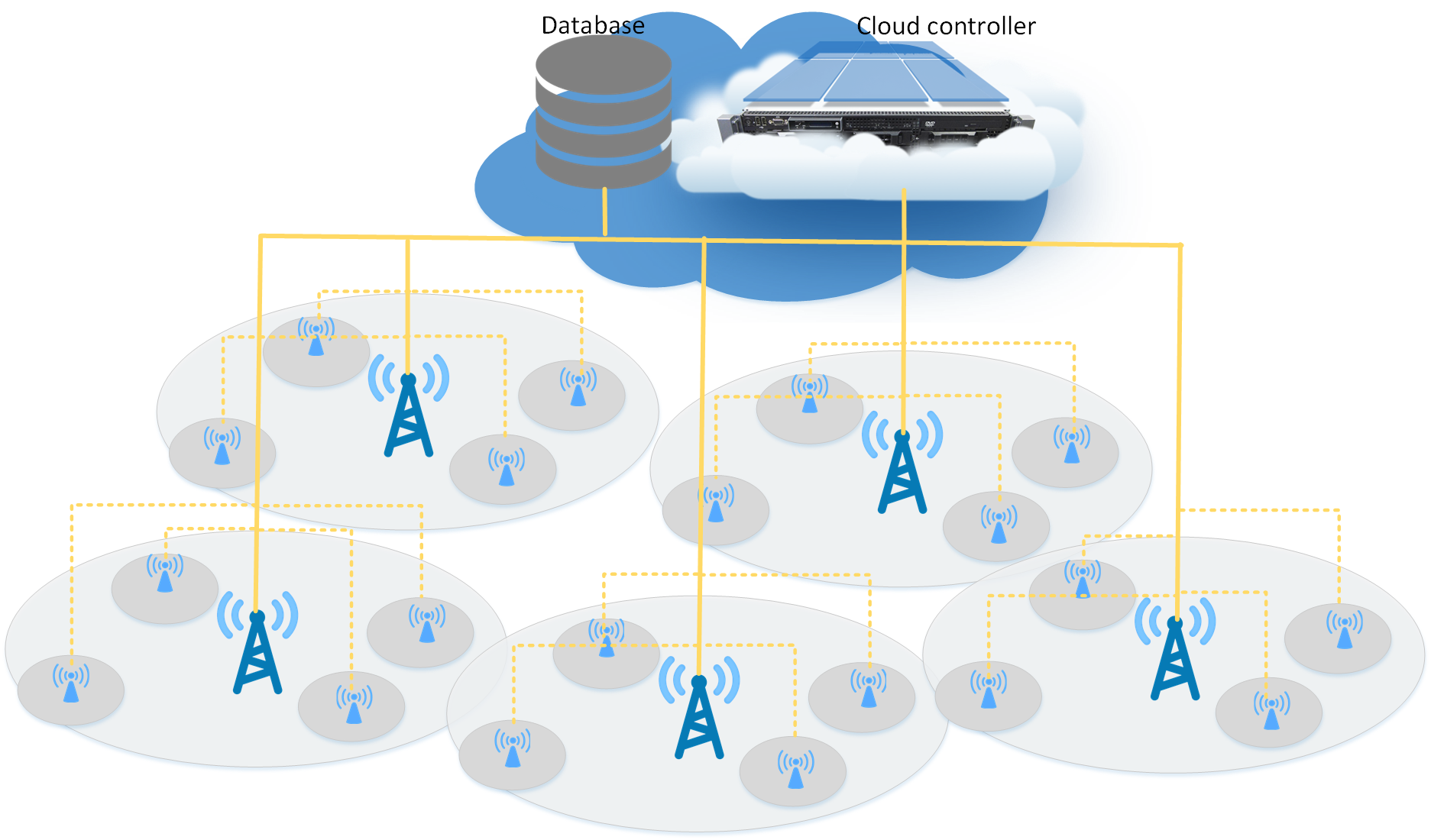}
	\caption{The cloud managed HetNet with a controller connecting to multiple macro and low-power small BSs.}
	\label{fig:Environment}
\end{figure}

\subsection{Mobile Traffic Offloading in HetNet}

The HetNet environment is a cellular network composed of cloud-connected macro and small BSs providing mobile transmission coverage over a geographical service region. The service region is divided into non-overlapping macrocell set $\mathcal{C}=\{1,\dots,C\}$, each with one macro BS deployed. Therefore, the set of macrocells fully covers the service region. Also, there is a set of overlapping small cells $\mathcal{K}^c=\{1,\dots,K^c\}$ in each macrocell $c\in\mathcal{C}$. A small cell contains a small BS to be turned ON (active) or OFF (inactive) to offload the traffic from macrocell $c$ dynamically. Every $T_r$ seconds, the mobile network collects the record of Internet activities $\mathbf{x}^c_n=(x^{m_c}_n,x^{s_c}_{n,1},\dots,x^{s_c}_{n,K})$ of all macrocells with discrete data \emph{recording time} index $n$ and stored in the database attached to the core network. The cloud controller dynamically adjusts the number of active small BSs in a macrocell every $M$ data records, so the discrete \emph{decision time} index $t$ is defined with decision period $T_d=M \cdot T_r$. Also, we define $y_{n,k}\in\{0,1\}$ as the indicator equals to 1 if the small BS $k$ in macrocell $c$ is ON and to 0 otherwise.
The number of active small BSs of macrocell $c$ at recording time $n$ is
\begin{equation}\label{eq:y}
    y^c_n = \sum_{k\in\mathcal{K}^c}y_{n,k}.
\end{equation}
Traffic loads of BSs within a macrocell are coupled with each other and distributed among macro and active small BSs based on a non-linear load coupling function $\mathbf{x}^c_n=\phi(\mathbf{x}^c_n)$. By applying a representative utility function, such as the logarithmic function, the cell load $\mathbf{x}^c_n$ can be evaluated\cite{Ho2014}. The proposed offloading architecture can adapt to different load distribution algorithms.

\subsection{Cell Load and Energy Efficiency}

The energy efficiency of the system can be assessed, given the recorded Internet activity of all cells. Under the mobile traffic demand, $D^c_n$, the data throughput of macrocell $c$ containing one macro BS and $K$ small BSs at recording time $n$ can be presented as:
\begin{equation}\label{eq:data_throughput}
  d^c_n = x^{m_c}_n + \sum_{k\in\mathcal{K}^c} x^{s_c}_{n,k}.
\end{equation}
Therefore, the property $d^c_n \leq D^c_n$ always holds.
The loading rates of macro and small BSs in macrocell $c$ can be written as:
\begin{align}\label{eq:bs_loading_rate}
  \rho^{m_c}_n &= \frac{x^{m_c}_n}{B^m \cdot T_r} \\
  \rho^{s_c}_{n,k} &= \frac{x^{s_c}_{n,k}}{B^s \cdot T_r},
\end{align}
where $B^m$ and $B^s$ are average achievable data rates when a user is served \emph{alone} by a macro or a small BS, respectively. For the whole macrocell, the traffic \emph{demand} rate over macro BS is:
\begin{equation}\label{eq:cell_loading_rate}
  \rho^c_n=\frac{D^c_n}{B^m \cdot T_r}.
\end{equation}

With cell loading rates, energy consumption during a recording period can be derived\cite{Saker2012}. For a macro BS, the energy consumption is:
\begin{equation}\label{eq:macro_energy_consumption}
e^{m_c}_n = P^m_{cst} + \alpha^m \rho^{m_c}_n P^m_{tx},
\end{equation}
where $P^m_{cst}$ is the constant power consumption caused by the signal processing unit, $P^m_{tx}$ is the transmission power, and $\alpha^m$ is the linear transmission power dependence factor. For a small BS in $k\in\mathcal{K}^c$, the energy consumption is:
\begin{equation}\label{eq:small_energy_consumption}
  e^{s_c}_{n,k} = [P^s_{cst} + \alpha^s\rho^{s_c}_{n,k}P^s_{tx}]\cdot y^c_{n,k}.
\end{equation}
$P^s_{cst}$, $P^s_{tx}$, and $\alpha^s$ are the constant signal processing power consumption, transmission power, and linear transmission power dependence factor, respectively. Also, we assume the small BS consumes power only when it is ON, i.e., $y^c_{n,k}=1$. Therefore, the total energy consumption of macrocell $c$ during recording time $n$ is:
\begin{equation}
  e^c_n = e^{m_c}_n + \sum^K_{k=1} e^{s_c}_{n,k}.
\end{equation}
The energy efficiency of macrocell $c$ at time step $n$ can thus be defined as the number of bits transmitted per Joule of energy as:
\begin{equation}\label{eq:EE}
  E^c_n = \frac{d^c_n}{e^c_n}.
\end{equation}

%
\section{DQN for Mobile Traffic Offloading}\label{sec:offloading}

The proposed DQN-based solution takes offloading action every decision time $t$. According to the pre-processed current state of a macrocell, $\psi(\mathbf{s}_t)$, the number of active small BSs in the macrocell, $a_t$, is determined to improve the reward $r_t$. Thus the Markov decision process (MDP) tuple is:
\begin{equation}\label{eq:mdp_tuple}
(\psi(\mathbf{s}_t), a_t , r_t, \mathbf{s}_{t+1}),
\end{equation}
where $\psi(\cdot)$ is the pre-processing function. Unique designs in the proposed offloading model include 1) \emph{statistics} of raw Internet activities as states to cope with unmatched recording and decision time scales; 2) traffic demand forecasting as pre-processing to improve offloading performance.

\subsection{The DQN Architecture}

\begin{figure}
\centering
\includegraphics[width=0.48\textwidth]{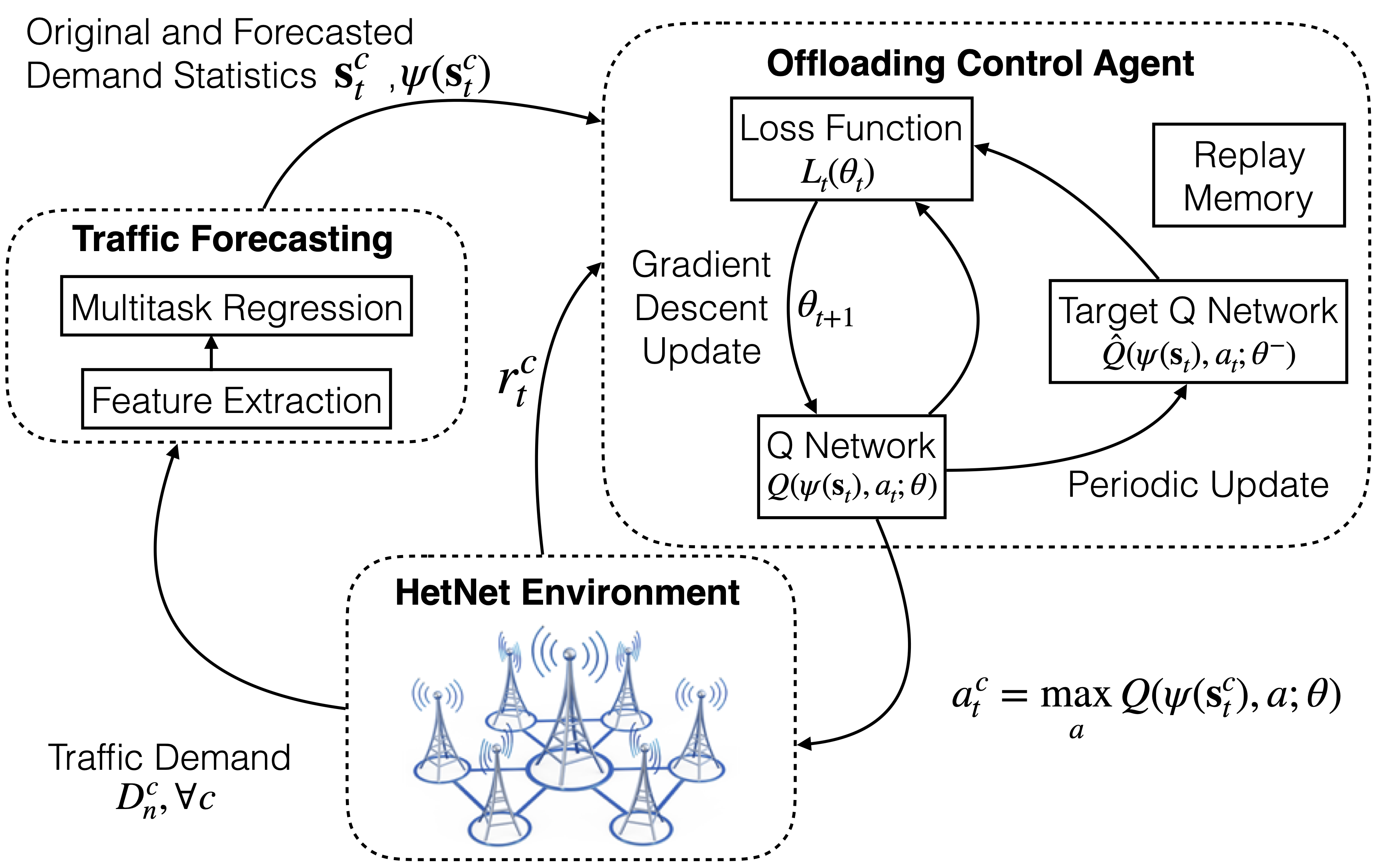}
\caption{The DQN architecture consists of the HetNet environment, traffic forecasting model, and control agent. Interactions between components are illustrated accordingly.}
\label{fig:DQN_arch}
\end{figure}

Figure~\ref{fig:DQN_arch} illustrates the interaction between the HetNet environment and a control agent in the cloud controller. The agent makes offloading decisions according to forecasted states. Internet activity records are stored in the replay memory and used for training. In this case, the mobile network traffic with continuous values is the inputs to the RL agent. It is challenging for the conventional Q-learning to construct Q tables with reasonable sizes. DQN parameterizes the Q-function and approximates it by a neural network to be more suitable for this application\cite{mnih2015human}. There are two neural networks in the DQN model \emph{shared by all macrocells}: the Q network is responsible for taking actions, and the target Q network operates only in the training stage. The agent determines ON/OFF status of small BSs according to the action learned from DQN.


The traffic forecasting module serves as the pre-processing function responsible for predicting the traffic load statistics through raw data collected from the environment. We develop a multitask learning structure for the module, which is designed specifically for network control data forecasting\cite{Huang2017}. The structure contains feature extraction and multitask regression stages. The feature extraction stage is a deep neural network integrating both CNN and RNN to catch both spatial and temporal traffic patterns. The forecasting module is further detailed in Section~\ref{ssec:forecasting}.

\subsection{Macrocell-Scale MDP Model}

The MDP model covers a macrocell instead of the whole network for better training efficiency. All macrocells share the same trained Q function. The \emph{state} at decision time $t$ is the statistics of mobile traffic \emph{demand} across an $N$-record \emph{observation window}. Thus the observation period, which can be different from the decision period $M$, is $T_o = N \cdot T_r$. For macrocell $c$, the minimum, average, and maximum of traffic demand at decision time $t$ are expressed as:
\begin{equation}
  \mathbf{s}^c_t = (D^c_{min,t}, D^c_{avg,t}, D^c_{max,t}),
\end{equation}
where
\begin{align}
  D^c_{\min,t} &= \min\{D^c_{tM-N+1}, D^c_{tM-N+2},\dots,D^c_{tM}\} \\
  D^c_{\mathrm{avg},t} &= \frac{1}{N}\sum^{tM}_{n=tM-N+1}D^c_n \label{eq:demand_avg}\\
  D^c_{\max,t} &= \max\{D^c_{tM-N+1}, D^c_{tM-N+2},\dots,D^c_{tM}\}.
\end{align}
Assuming the pre-processing function predicts the statistics in the coming decision period, the current state becomes
\begin{equation}
\psi(\mathbf{s}^c_t) = (\hat{D}^c_{min,t+1}, \hat{D}^c_{avg,t+1}, \hat{D}^c_{max,t+1}).
\end{equation}
The \emph{action}, $a_t$, at decision time $t$ is the number of active small BSs in a macrocell. The ON/OFF status of small cells remains unchanged during a decision window, i.e., $a^c_t = y^c_{tM+1} = y^c_{tM+2} = \dots = y^c_{(t+1)M}$.

The \emph{reward} at decision time $t$ of macrocell $c$ is labeled as $r^c_t$ evaluating the return of actions considering both energy consumption and traffic loading. $r^c_t$ is calculated as:
\begin{equation}\label{eq:total_reward}
  r^c_t = \beta \cdot r^c_{E,t} + \gamma \cdot r^c_{\rho,t},
\end{equation}
where $r^c_{E,t}$ and $r^c_{\rho,t}$ are the reward components of energy efficiency and BS loading rates, respectively. $\beta, \gamma$ are scaling factors. The energy efficiency component is defined as the average over a decision period:
\begin{equation}\label{eq:reward_componet_of_ee}
  r^c_{E,t} = \frac{1}{M}\sum^{(t+1)M}_{n=tM+1}E^c_n.
\end{equation}
The reward component is designed to punish QoS degradation due to high traffic loads and encourage active small BSs running under desired loading rates. $r^c_{\rho,t}$ can be computed by:
\begin{equation}\label{eq:reward_componet_of_cell_loading}
\begin{aligned}
r^c_{\rho,t} =&
  \begin{cases}
  -\exp(w_1(\rho^{m_c}_t-\rho_{th_1})) & \text{if $\rho^{m_c}_t>\rho_{th1}$} \\
  0                                   & \text{otherwise}
  \end{cases} \\
+  & \sum_{k \in K^c}
  \begin{cases}
  -\exp(w_2(\rho_{th1}-\rho^{s_c}_{t,k})) & \text{if } \rho^{m_c}_t<\rho_{th1} \land \\                         & \quad \rho^{s_c}_{t,k}<\rho_{th2} \\
  \exp(w_3(\rho^{s_c}_{t,k})) & \text{if $\rho_{th1}>\rho^{s_c}_{t,k}>\rho_{th2}$} \\
  -\exp(w_4(\rho^{s_c}_{t,k}-\rho_{th1})) & \text{if $\rho^{s_c}_{t,k}>\rho_{th1}$} \\
  0 & \text{otherwise},
  \end{cases}
\end{aligned}
\end{equation}
where
\begin{align}
  \rho^{m_c}_t &= \frac{1}{M}\sum^{(t+1)M}_{n=tM+1}\rho^{m_c}_n \\
  \rho^{s_c}_{t,k} &= \frac{1}{M}\sum^{(t+1)M}_{n=tM+1}\rho^{s_c}_{n,k}.
\end{align}
$\rho_{th1}$ and $\rho_{th2}$ are pre-determined loading rate thresholds with $\rho_{th1}>\rho_{th2}$. It is used to set the desired operation region. The first term in~\eqref{eq:reward_componet_of_cell_loading} is related to the loading rate of macro BS $m_c$. The design is to punish $r^c_{\rho,t}$ when it goes beyond the loading threshold $\rho_{th1}$ by exponential increment. The second term in \eqref{eq:reward_componet_of_cell_loading} is related to the loading rate of small BS $k$ with three conditions. The first condition punishes $r^c_{\rho,t}$ only when the macro BS loading rate is lower than $\rho_{th1}$, and small BS loading rate $\rho^{s_c}_{t,k}$ is lower than $\rho_{th2}$. The condition prevents turning on too many small BSs. The second condition increases $r^c_{\rho,t}$ when loading rates of small BSs $\rho^{s_c}_{t,k}$ is between $\rho_{th1}$ and $\rho_{th2}$, which encourages operating at proper loading rates. The third condition is also a punishment that decreases $r^c_{\rho,t}$ when $\rho^{s_c}_{t,k}$ exceeds $\rho_{th1}$. $w_1, w_2, w_3, w_4$ are adjustable weights.



\subsection{DQN Training}\label{ssec:training}

Two mechanisms that make DQN work. First, DQN uses experience replay as the most important trick. During the training process, experiences as MDP tuples are stored in the replay memory shown in Fig.~\ref{fig:DQN_arch} and are randomly sampled instead of using the most recent transition. Experience replay breaks the similarity of the continuous training sample which otherwise might cause the network into a local minimum \cite{lin1993reinforcement}. Second, DQN maintains two separate Q-networks. $Q(\psi(\mathbf{s}_t),a_t;\theta)$ has current parameters $\theta$ while $\hat{Q}(\psi(\mathbf{s}_t),a_t;\theta^-)$ has old parameters $\theta^{-}$ called target Q-network. The current parameters $\theta$ are updated per time-step during the training stage. By contrast, the old parameters $\theta^-$ are updated only after several iterations. Based on the loss function $L_t(\theta_t)$ of each iteration, the parameter update is executed as:
\begin{align}\label{eq:theta_update}
  \theta_{t+1} &= \theta_t - \frac{1}{2}\eta\nabla_{\theta_t}L_t(\theta_t) \nonumber\\
    &= \theta_t - \frac{1}{2}\eta\nabla_{\theta_t}\left[r_t+\mu\max_a \hat{Q}(\psi(\mathbf{s}_{t+1}),a;\theta^-) \right.\nonumber\\
    &\qquad\left.\vphantom{\max_a} -Q(\psi(\mathbf{s}_t),a_t;\theta_t)\right]^2 \nonumber\\
    &=\theta_t + \eta\left[r_t+\mu\max_a \hat{Q}(\psi(\mathbf{s}_{t+1}),a;\theta^-) \right.\nonumber\\
    &\qquad\left.\vphantom{\max_a} -Q(\psi(\mathbf{s}_t),a_t;\theta_t)\right]\nabla_{\theta_t}Q(\psi(\mathbf{s}_t),a_t;\theta_t).
\end{align}
The action for a specific macrocell $c$ is then determined by:
\begin{equation}\label{eq:act}
  a^c_t = \arg\max_a Q(\psi(\mathbf{s}^c_t),a;\theta).
\end{equation}
By applying \eqref{eq:act} to each macrocell, efficient offloading across the whole map area can be achieved.





%
\section{Data Processing and Forecasting}\label{sec:data}

The public telecommunication dataset and mobile traffic forecasting model applied for our data-driven training and experiments are introduced in this section.

\subsection{Mobile Internet Traffic Data}\label{ssec:traffic_data}

For a data-driven study on mobile networks, a high-quality dataset is essential and often challenging to obtain. In this work, one of the most comprehensive public datasets released by Telecom Italia in 2015 is applied\cite{Barlacchi2015}. It was initially created for a big data challenge with projects ranging from mobile networking to social applications. Provided data points include records in telecommunication, weather, news, social activity, and electricity during November and December 2013. Internet traffic activity data from the city of Milan is used explicitly in this work.

\begin{figure}
\centering
\subfloat[Milan grid traffic heat map. The blue square denotes the downtown area.]{\includegraphics[width=0.35\textwidth]{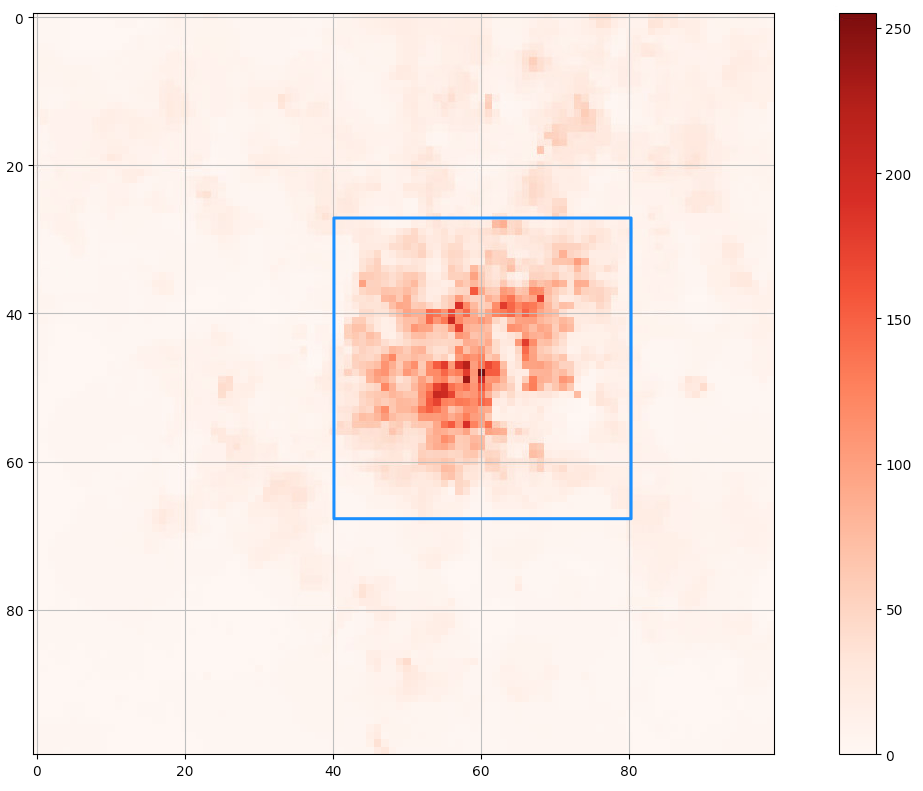}
\label{fig:Milan_grid_heat_map}}
\hfil
\subfloat[Macrocell coverage in downtown Milan. The red dots denotes macro BSs.]{\includegraphics[width=0.4\textwidth]{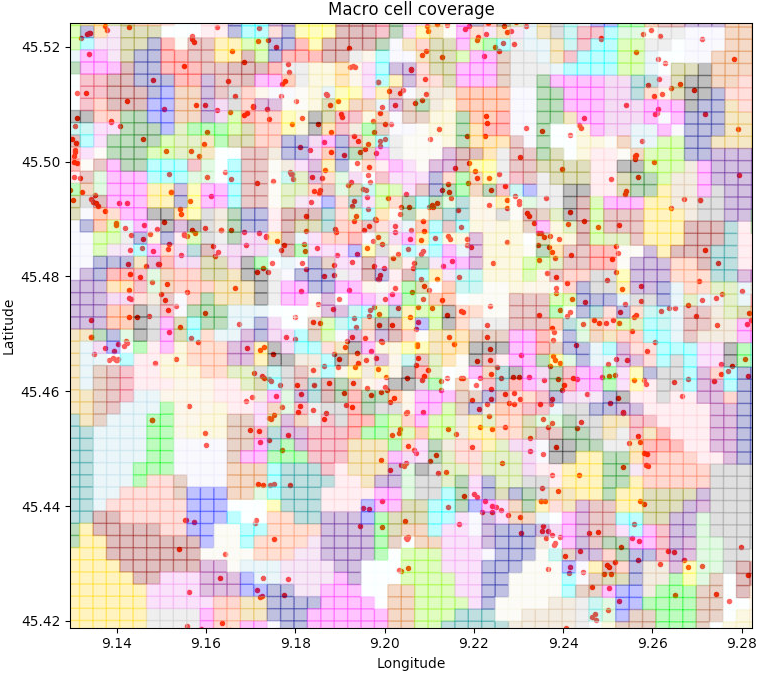}
\label{fig:macro_cell_coverage}}
\caption{The traffic heat map and macrocell coverage on Milan grids.}
\label{fig:cell_maps}
\end{figure}

Milan city is divided into a set of $100 \times 100$ geographical grids, $\mathcal{I}=\{1,\dots,I\}$. Each grid has a unique square ID covering an area of $235 \times 235$ square meters. The data is described with standard WGS84 (EPSG:4326) GeoJson format. \figurename~\ref{fig:Milan_grid_heat_map} illustrates the traffic heat map of the whole city. A $12\times12$ high loading sub-area of downtown is considered in the experiments. Part of the telecommunication data are the following:
\begin{itemize}
	\item Square ID: the identification of squares in the Milan grids.
	\item Time interval: the beginning time of a data recording interval. The recording interval is 10-minute long for each record, i.e., $T_r=600$.
	\item Internet traffic activity: the aggregated number of data transmission activities generated during the time interval in this grid. By multiplying a value of average data size per activity, we can restore the amount of transmitted data.
\end{itemize}
For each grid, $i\in\mathcal{I}$, the aggregated amount of traffic, $g_{n,i}$, is collected every recording time $n$ and utilized as traffic demand in the data-driven experiments. The Internet traffic activity records are referred to as records in later sections. Nine-tenths of records are allocated for training. The remaining records (about six days or 149 hours long) are the testing set.

\subsection{Base Station Data}\label{ssec:bs_data}


Since the Telecom Italia dataset contains no BS information, public data from OpenCellID\cite{OpenCellID} are adopted to implement the network environment. OpenCellID contains the following information for each BS (cell tower):
\begin{itemize}
	\item Type: the type of BS in GSM, UMTS, or LTE.
	\item MCC: the mobile country code of where the cell tower located.
	\item MNC: the mobile network code as the unique identification of a mobile network operator.
	\item Longitude: the longitude coordinate of the cell tower.
	\item Latitude: the latitude coordinate of the cell tower.
\end{itemize}
We extract the BS information for Telecom Italic through MCC and MNC. Based on the coordinates, cell towers are mapped on to the Milan grid to set locations of \emph{macro} BSs, as red dots shown in \figurename~\ref{fig:macro_cell_coverage}. The adjacent grids are further grouped into macrocells to define the coverage of a macro BS. We assume that a grid is covered by the closest macro BS according to the Euclidean distance between the macro BS and grid centers. The set of geographical grids covered by a macrocell $c$ is denoted as $\mathcal{I}^c=\{1,\dots,I^c\}$. Therefore, the traffic demand of a macrocell $c$ at recording time $n$ is:
\begin{equation}\label{eq:traffic_demand}
D^c_n = \sum_{i\in\mathcal{I}^c}g_{n,i}
\end{equation}
For each macrocell, we assume there are $K^c$ small BSs in it.



%
%
\subsection{Mobile Traffic Forecasting}\label{ssec:forecasting}

The traffic forecasting model utilizes a deep learning-based structure explicitly designed for mobile Internet management\cite{Huang2017}. In this work, the model predicts the maximum, average, and minimum traffic demands in the coming hour based on data in the previous hour. That is, the model takes the traffic demand, $D^c_n$, across the observation window as inputs and produces forecasted statistics, $\psi(\mathbf{s}^c_t)$, as outputs. Also, in the experiments, the observation and decision windows are identical, $N=M=6$. The neural network adopts a structure combining CNN, RNN, and multitask learning to take advantage of spatial, temporal, and cross-task patterns. The adopted model significantly improves the prediction of maximum traffic demand in the next decision window over traditional methods.
\section{Performance Evaluation}\label{sec:results}

In this section, we detail the data-driven training environment built upon the Telecom Italia dataset\cite{Barlacchi2015} with performance results compared with major offloading approaches.

\subsection{Data-Driven HetNet Simulation Setup}

We construct the environment based on public mobile traffic records and BS information. The CNN-RNN traffic forecasting model with multitask learning\cite{Huang2017} is also used due to its balanced prediction performance. There are ten small cells ($K=10$) to be controlled by the cloud controller in each macrocell. The Q network and target Q network have identical structures having three-layer and 512-512-11 neural nodes in each layer. The output layer has a $K+1$ number of nodes to indicate actions activating zero to $K$ small BSs. A full list of parameters and values is in Table~\ref{table:offloading_parameter}.

\begin{table}
\renewcommand{\arraystretch}{1.3}
\caption{Simulaiton Parameters\cite{chen2015energy,Saker2012}}
\label{table:offloading_parameter}
\centering
\begin{tabular}{|c|c|c|}
\hline
  Parameter & Value \\ \hline\hline
  \multicolumn{2}{|c|}{Environment} \\ \hline
  $B^m$, $B^s$ & 12.5, 27 Mb/s \\ \hline 
  $P^m_{cst}$, $P^s_{cst}$ & 130, 4.8 W \\ \hline
  $P^m_{tx}$, $P^s_{tx}$ & 20, 1 W \\ \hline
  $\alpha^m$, $\alpha^s$ & 4.7, 8 \\ \hline
  $\beta$, $\gamma$ & 100, 1 \\ \hline
  $\rho_{th1}$, $\rho_{th2}$ & 0.7, 0.5 \\ \hline
  $w_1$, $w_2$, $w_3$, $w_4$ & 10 \\ \hline
  Data recording period, $T_r$ & 600 s \\ \hline
  Total Number of grids, $I$ & 144 \\ \hline
  Number of small cells in a macrocell, $K$ & 10 \\ \hline
  Observation and decision window size, $N=M$ & 6 \\ \hline
  Avg. data size per activity & 15 Mb \\ \hline
  \multicolumn{2}{|c|}{Deep Q Networks} \\ \hline
  Number of layers & 3 \\ \hline
  Number of nodes in each layer & 512:512:11 \\ \hline
  Learning rate, $\mu$ & 0.001 \\ \hline
  Discount factor, $\eta$ & 0.9 \\ \hline
  Exploration rate & 0.1 \\ \hline
  Iterations per target network update & 200 \\ \hline
\end{tabular}
\end{table}

The performance of our offloading model is compared with the other four schemes and explained below.
\begin{itemize}
  \item Macrocell only (MACRO): only the macrocell BS in each cell $c$ is transmitting. There is no small BS for traffic offloading.
  \item Static offloading (STATIC): the macro and all $K$ small BSs are turned ON.
  \item Q-learning offloading\cite{chen2015energy} (QL): the small BS makes ON/OFF decisions based the on Q-table RL approach. To reduce the number of states and maintain a reasonable Q-table size, we quantize the raw Internet traffic value with a step size of 20 activities. There is no deep network applied in this method.
  \item DQN Offloading (DQN): the proposed DQN offloading model is applied in this method without traffic forecasting. That is, the MDP tuple is $(\mathbf{s}^c_t, a^c_t, r^c_t, \mathbf{s}^c_{t+1})$.
  \item DQN offloading with traffic forecasting (DQN-F): the complete proposed method. The forecasting function $\psi(\mathbf{s}^c_t)$ is used as states in DQN offloading.
\end{itemize}

To further analyze the results, we classify 144 geological grids into five groups according to macrocell traffic demand rates, i.e., demand groups. According to the average traffic demand, the demand rate of macrocell $c$ at decision time $t$ is:
\begin{equation}\label{eq:demand_rate}
  \rho^{D_c}_t=\frac{D^c_{avg,t}}{B^m\cdot T_r},
\end{equation}
where $D^c_{avg,t}$ can be evaluated from the dataset by \eqref{eq:demand_avg} and \eqref{eq:traffic_demand}. A demand rate greater than 0.7 is considered high. A rate greater than 1 indicates overloading. Table~\ref{table:group_location_by_macro_load} presents group profiles containing numbers of grids, average traffic demand rate, and the average amount of traffic.

\begin{table*}
\renewcommand{\arraystretch}{1.3}
\centering
\caption{Geographical Grids Groups by Macrocell Traffic Demand Rates}
\label{table:group_location_by_macro_load}
\begin{tabular}{|c|c|c|c|c|c|}
\hline
  Macrocell traffic demand rate ($\rho^{D_c}_t$) group & $0\sim0.3$ & 0.3$\sim$0.7 & 0.7$\sim$1 & 1$\sim$2 & $>2$ \\ \hline\hline
  Number of geographical grid & 19 & 54 & 15 & 42 & 14 \\ \hline
  Avg. traffic demand rate & 0.16 & 0.49 & 0.84 & 1.37 & 2.64 \\ \hline
  Avg. amount of traffic per grid (Gb) & 1092.29 & 3326.54 & 5682.62 & 9197.46 & 17747.24  \\ \hline
\end{tabular}
\end{table*}

\subsection{Energy Efficiency}

The overall energy efficiency results are shown in Table~\ref{table:each_method_overview}. The values are evaluated using 149 hours of testing data and averaged across macrocells. The statistics include total energy consumption, transmitted data, and energy efficiency. The proposed DQN-F scheme outperforms the other four for 88\%, 16\%, 8.5\%, and 7\% in energy efficiency, respectively. The outcome reveals that the DQN offloading architecture can effectively adjust the number of active small cells considering energy efficiency. Also, the advantage of DQN-F over DQN indicates that traffic forecasting is beneficial in the way of providing proactive decisions to future traffic demands.

\begin{table*}
\renewcommand{\arraystretch}{1.3}
\centering
\caption{Overall Energy Efficiency}
\label{table:each_method_overview}
\begin{tabular}{|l|c|c|c|}
\hline
Offloading Methods & Data Transmitted (Gb) & Energy Consumption (M-Joule) & Energy Efficiency (Mb/Joule) \\
\hline\hline
DQN with forecasting (DQN-F) & 6362.15 & \textbf{86.37} & \textbf{0.0737} \\ \hline
DQN Offloading (DQN) & 6200.12 & 89.95 & 0.0689 \\ \hline
Q-learning offloading (QL) & 6121.06 & 90.14 & 0.0679 \\ \hline
Static offloading (STATIC) & \textbf{6391.06} & 101.16 & 0.0631 \\ \hline
Macrocell only (MACRO)  & 3980.75 & 101.58 & 0.0391 \\ \hline
\end{tabular}
\end{table*}

\begin{figure}
\centering
\subfloat[Data Transmitted.]{\includegraphics[width=0.47\textwidth]{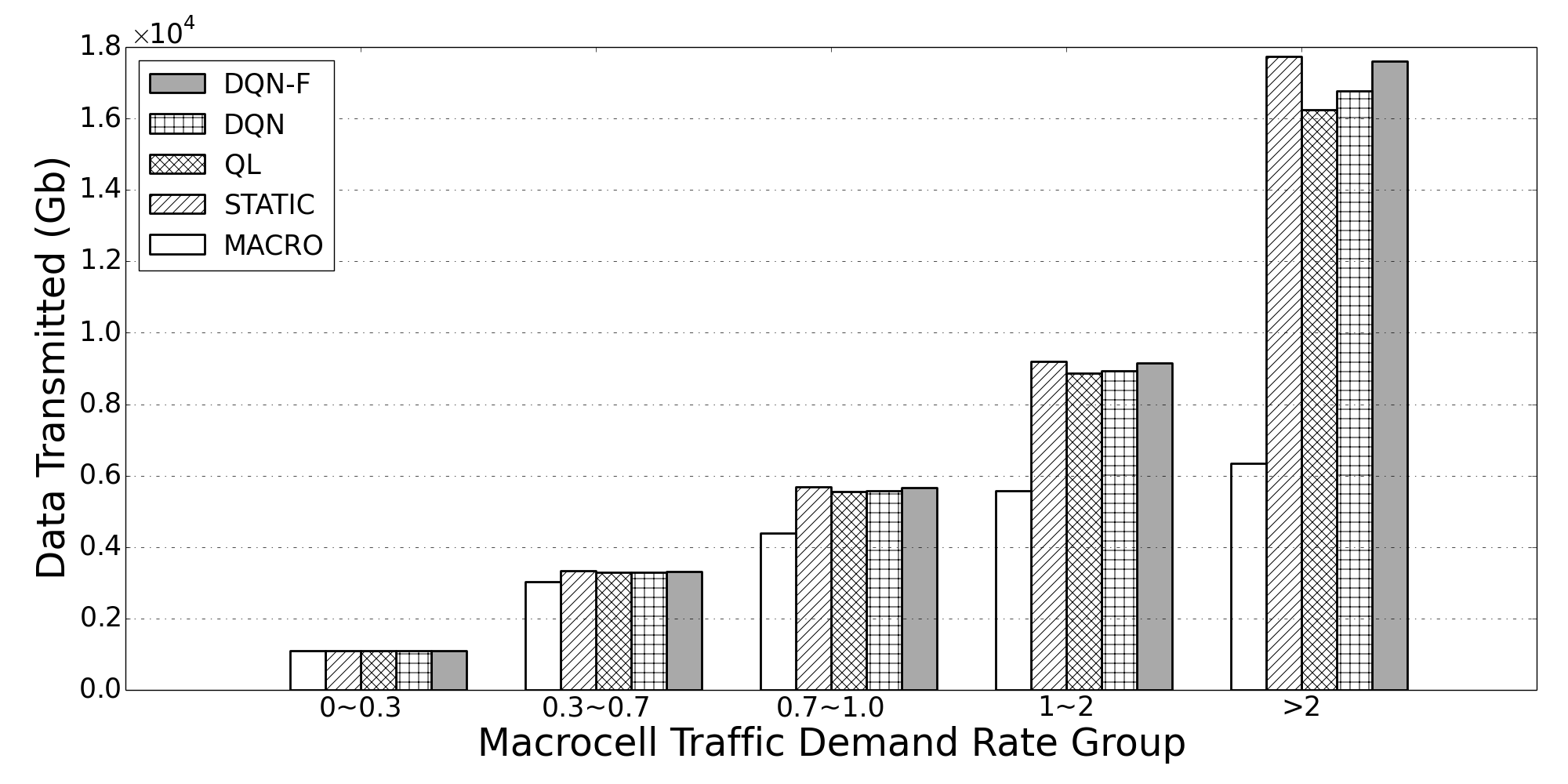}
\label{fig:transmitted_data}}
\hfil
\subfloat[Total Energy Consumption.]{\includegraphics[width=0.47\textwidth]{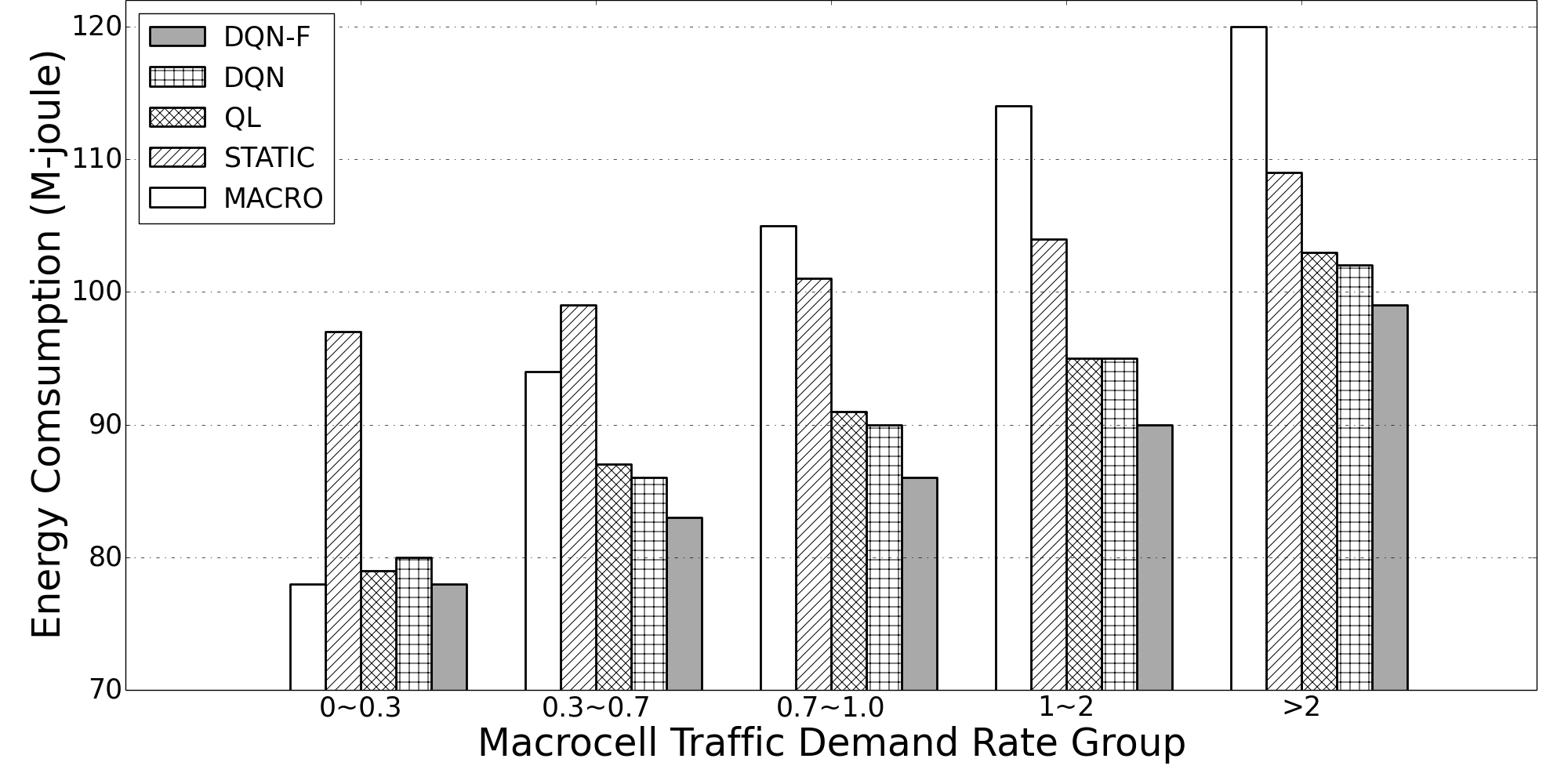}
\label{fig:total_energy_comsuption}}
\hfil
\subfloat[Energy Efficiency.]{\includegraphics[width=0.47\textwidth]{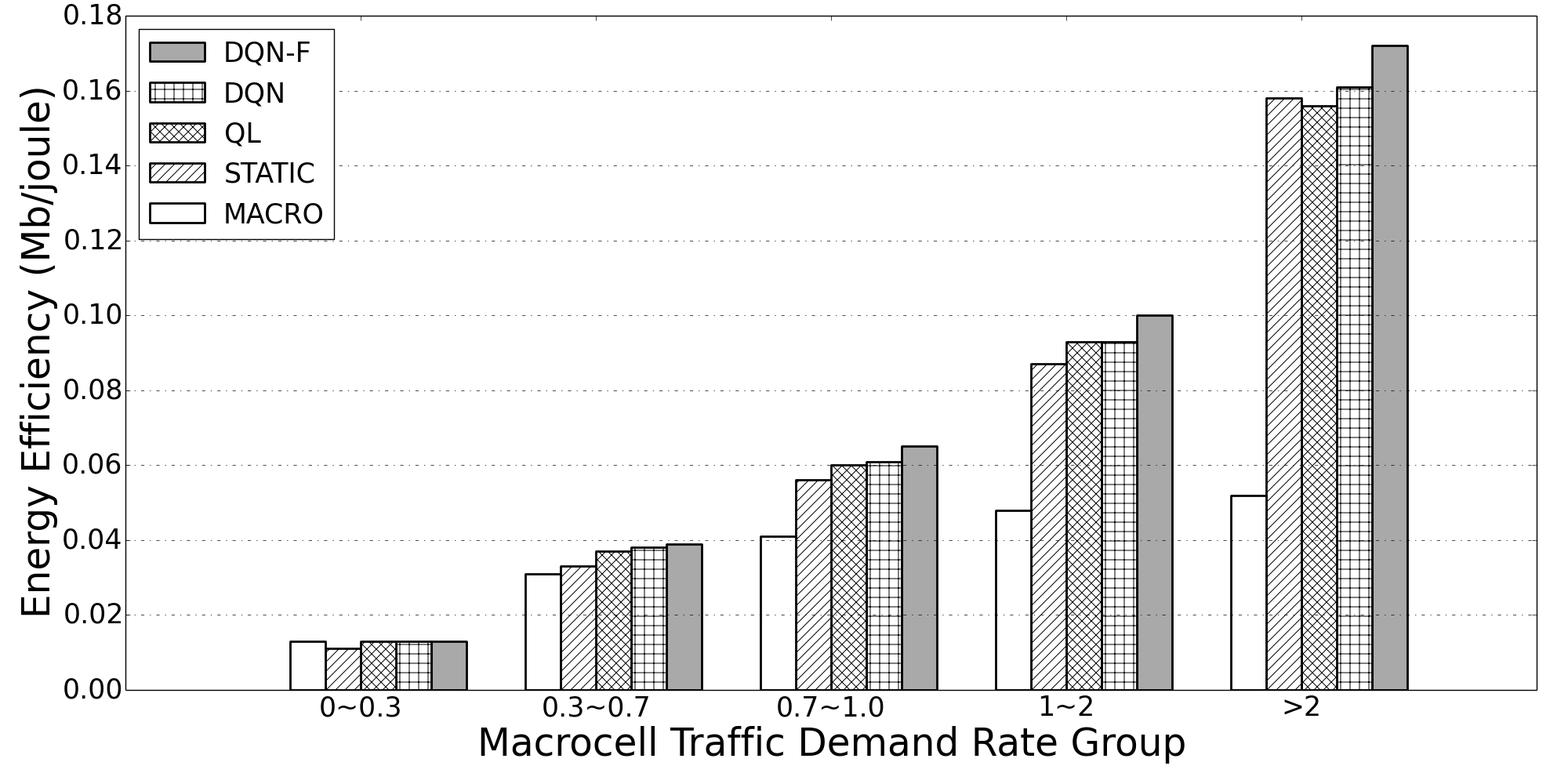}
\label{fig:energy_efficiency}}
\caption{Energy efficiency performance in orginal macrocell loading groups.}
\label{fig:energy_effiency_comparison}
\end{figure}

Fig.~\ref{fig:energy_effiency_comparison} details the energy efficiency performance in original macrocell loading groups. When the demand rate is below 0.7, a similar amount of data is transmitted by all schemes. The energy efficiency of the MACRO scheme remains comparable with others. The STATIC method consumes much more energy due to enabling more small cells. When the loading is close to 1, RL-based cases QL, DQN, and DQN-F perform better. The proposed DQN-F outperforms QL by 5\% and 8\% in 0.3$\sim$0.7 and 0.7$\sim$1 loading groups, respectively. Comparing DQN-F with DQN, the forecasting function shows 2\% and 6\% advantages in 0.3$\sim$0.7 and 0.7$\sim$1 loading groups.

In the loading group 1$\sim$2, the STATIC case represents a baseline energy efficiency when the small cells are always on. The DQN-F scheme transmitted almost the same amount of data with STATIC while consuming less energy. The difference between DQN-F and DQN in energy efficiency increases to 7.5\%. When the macrocells are severely overloaded in the group 2$\sim$10, the energy efficiency grows sharply because most of the data traffic demands are serviced by small cells. The proposed DQN-F performs the best with 5\% more data transmitted and 3\% less energy consumed than DQN. The energy efficiency is also 10\% better than QL, with 8.4\% more data transmitted.

\subsection{Loading Rate Distribution}

\begin{figure}
\centering
\subfloat[Macro BS loading rate after offloading.]{\includegraphics[width=0.47\textwidth]{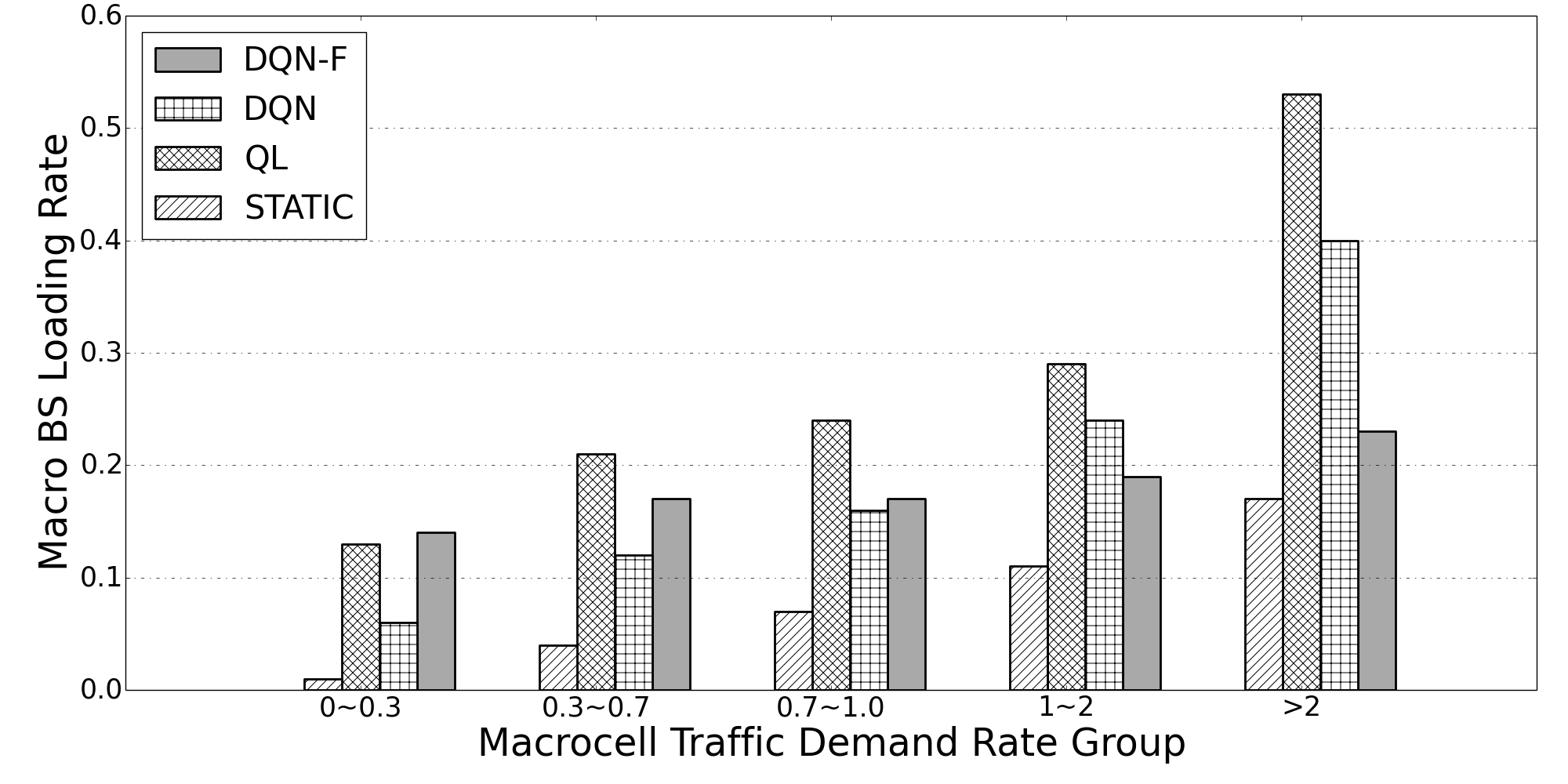}
\label{fig:macro_loading_rate_after}}
\hfil
\subfloat[Small BS loading rate.]{\includegraphics[width=0.47\textwidth]{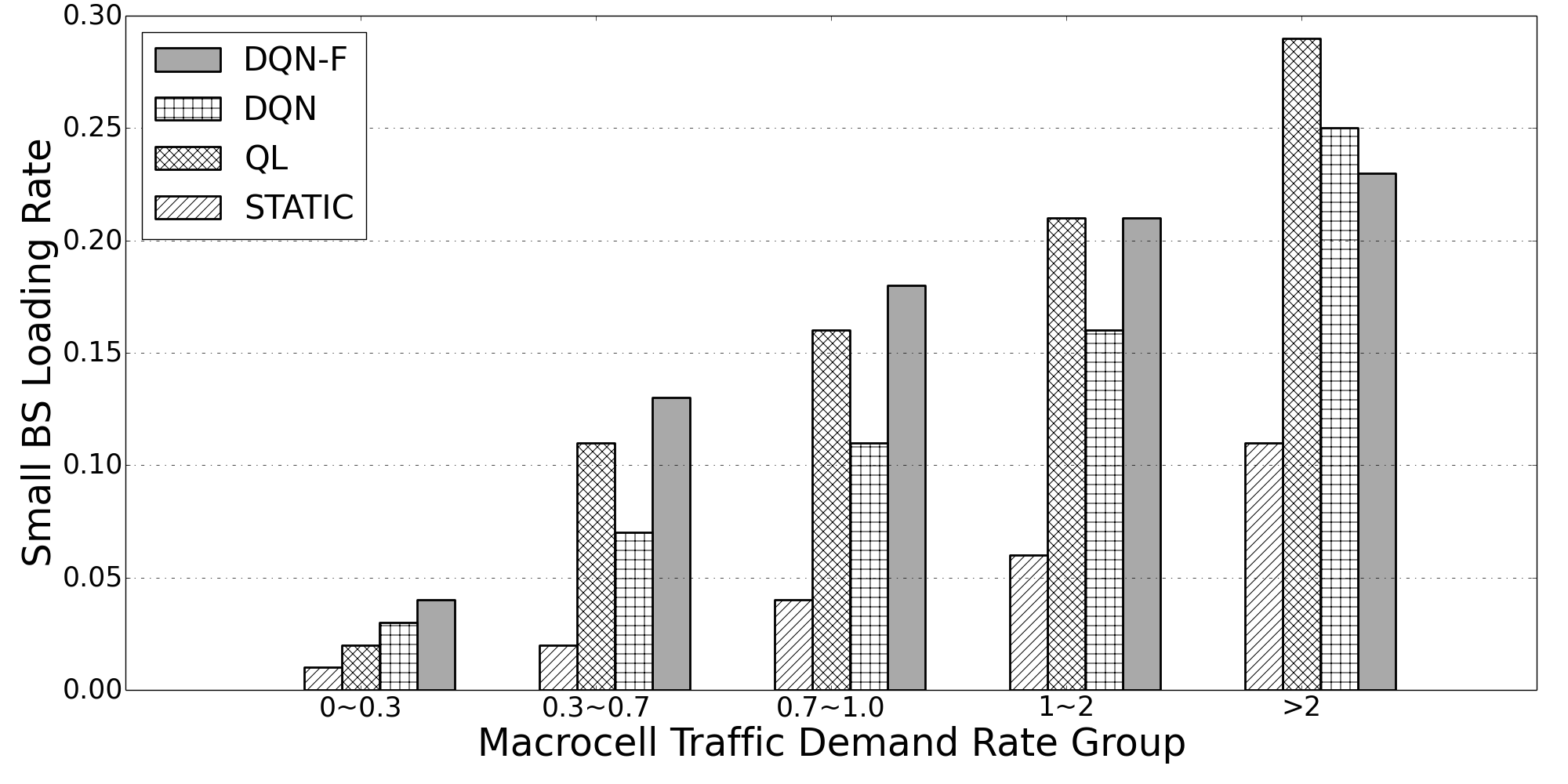}
\label{fig:small_loading_rate}}
\hfil
\subfloat[Number of active small BS in a macrocell.]{\includegraphics[width=0.47\textwidth]{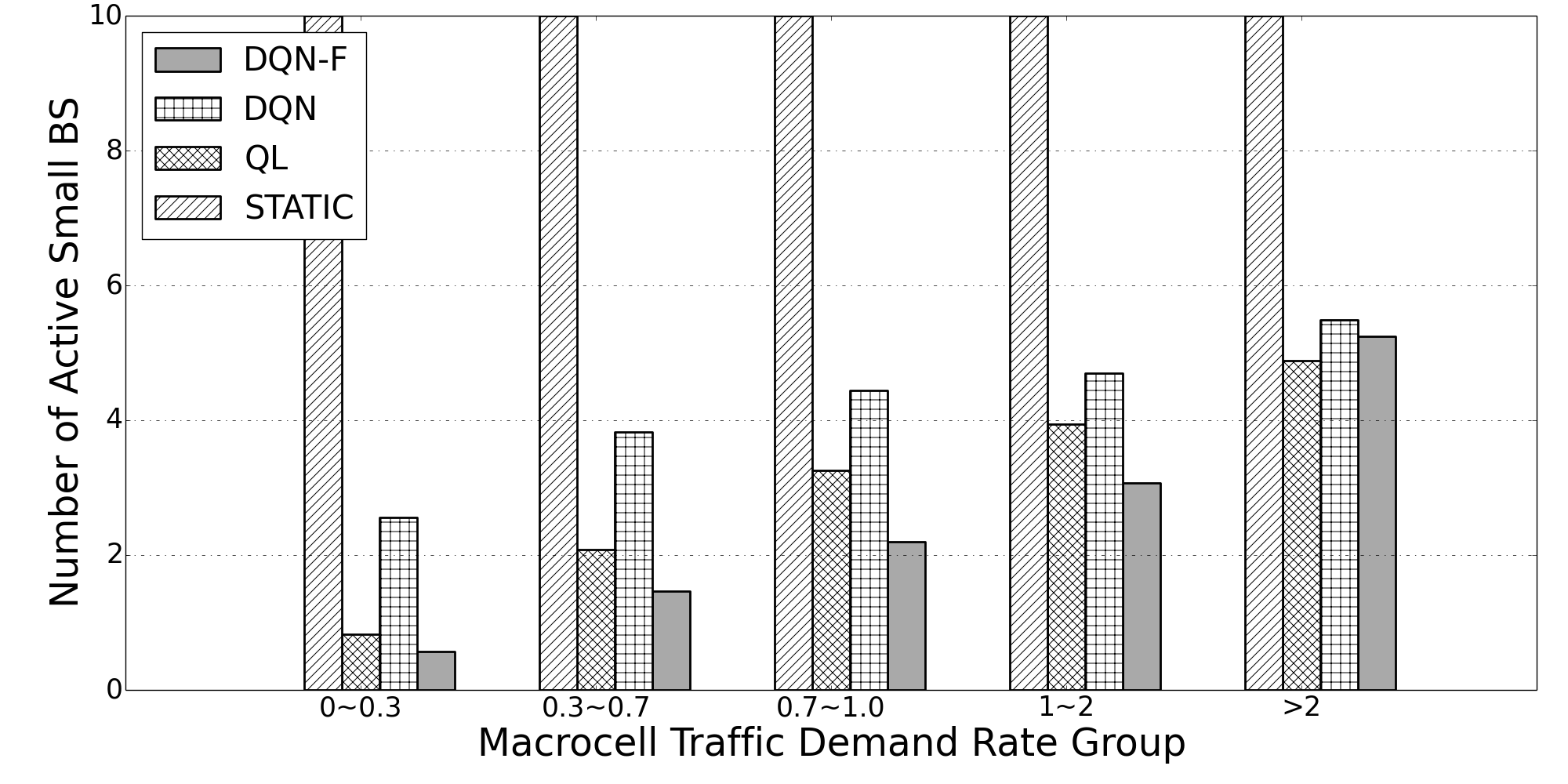}
\label{fig:small_active}}
\caption{BS loading rates and number of active small BSs after offloading in demand rate groups.}
\label{fig:cell_loading_rate_comparison}
\end{figure}

Figure~\ref{fig:cell_loading_rate_comparison} shows the BS loading rates and number of active small BSs after offloading in demand rate groups. For macrocells with demand rates between 0 to 2, all RL-based schemes, including QL, DQN, and DQN-F, provide low macro BS loading rates after offloading. DQN-F shows higher efficiency by fewer active small BSs working in preferred loading rates, which means traffic forecasting improves offloading decisions. When the demand rates are higher than 2, the average number of active small BSs are close to other RL-based methods, while QL and DQN resulting in higher macro and small BS loading rates than proposed DQN-F. That is, using a similar number of small BSs, DQN-F activates them in better locations and moments to efficiently offload the dynamic amount of traffic.

\begin{figure}
\centering
\subfloat[Resulting macro BS loading rate distribution for demand group $\rho^{D_c}_t=1\sim2$.]{\includegraphics[width=0.48\textwidth]{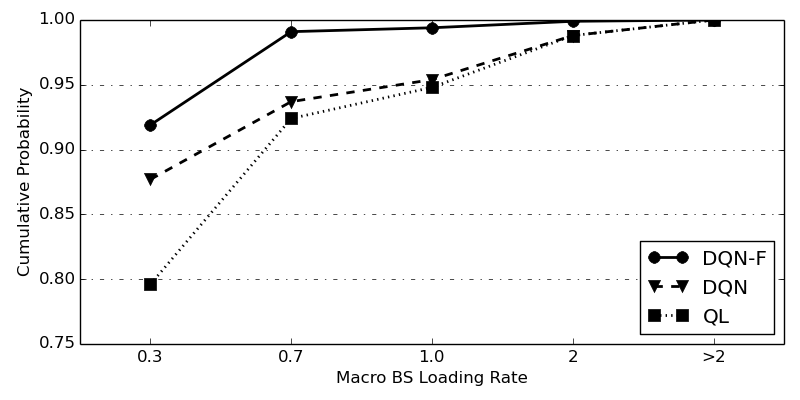}
\label{fig:macro_load_rate_after_offloading_group_4}}
\hfil
\subfloat[Resulting macro BS loading rate distribution for demand group $\rho^{D_c}_t>2$.]{\includegraphics[width=0.48\textwidth]{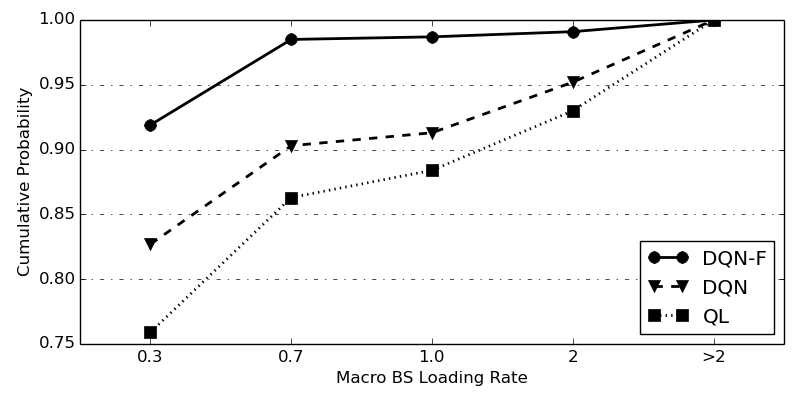}
\label{fig:macro_load_rate_after_offloading_group_5}}
\caption{Distribution of macro BS loading rates after offloading decisions for two overloaded demand groups.}
\label{fig:macro_load_rate_after_offloading}
\end{figure}

Figure~\ref{fig:macro_load_rate_after_offloading} illustrates the distribution of macro BS loading rates \emph{after} offloading decisions, $\rho^{m_c}_t$, of two overloaded demand groups, $\rho^{D_c}_t=1\sim2$ and $\rho^{D_c}_t>2$, in the cumulative probability of total decision periods. The QL has up to 5.2\% and 11.6\% of the decision considered failed with macro BS still overloaded. DQN-F, by contrast, has only 0.6\% and 1.3\% of decisions resulting in overloaded macro BS. DQN, the case without traffic forecasting, which has up to 4.6\% and 8.7\% failed decisions, showing the performance degradation when the traffic provision is not satisfactory.


%
\section{Conclusions}\label{sec:conclusion}

The article studies the energy-aware small BS management in HetNet. The deep reinforcement learning-based offloading architecture consists of a data-driven environment, a smart offloading control agent, and a traffic forecasting model. Combining DQN, CNN, and RNN techniques, the proposed control scheme outperforms offloading approaches using conventional Q-learning or without traffic prediction. It successfully demonstrates a novel usage of deep learning on SON management tasks.

%


\bibliographystyle{IEEEtran}
\bibliography{DRLOffloading}

\begin{thebibliography}{10}
\providecommand{\url}[1]{#1}
\csname url@samestyle\endcsname
\providecommand{\newblock}{\relax}
\providecommand{\bibinfo}[2]{#2}
\providecommand{\BIBentrySTDinterwordspacing}{\spaceskip=0pt\relax}
\providecommand{\BIBentryALTinterwordstretchfactor}{4}
\providecommand{\BIBentryALTinterwordspacing}{\spaceskip=\fontdimen2\font plus
\BIBentryALTinterwordstretchfactor\fontdimen3\font minus
  \fontdimen4\font\relax}
\providecommand{\BIBforeignlanguage}[2]{{%
\expandafter\ifx\csname l@#1\endcsname\relax
\typeout{** WARNING: IEEEtran.bst: No hyphenation pattern has been}%
\typeout{** loaded for the language `#1'. Using the pattern for}%
\typeout{** the default language instead.}%
\else
\language=\csname l@#1\endcsname
\fi
#2}}
\providecommand{\BIBdecl}{\relax}
\BIBdecl

\bibitem{aijaz2013survey}
A.~Aijaz, H.~Aghvami, and M.~Amani, ``A survey on mobile data offloading:
  technical and business perspectives,'' \emph{IEEE Wireless Communications},
  vol.~20, no.~2, pp. 104--112, 2013.

\bibitem{Zhang2015}
N.~Zhang, N.~Cheng, A.~T. Gamage, K.~Zhang, J.~W. Mark, and X.~Shen, ``{Cloud
  assisted HetNets toward 5G wireless networks},'' \emph{IEEE Communications
  Magazine}, vol.~53, no.~6, pp. 59--65, Jun. 2015.

\bibitem{Imran2014}
A.~Imran and A.~Zoha, ``{Challenges in 5G: how to empower SON with big data for
  enabling 5G},'' \emph{IEEE Network}, vol.~28, no.~6, pp. 27--33, Nov. 2014.

\bibitem{Han2016}
F.~Han, S.~Zhao, L.~Zhang, and J.~Wu, ``{Survey of Strategies for Switching Off
  Base Stations in Heterogeneous Networks for Greener 5G Systems},'' \emph{IEEE
  Access}, vol.~4, pp. 4959--4973, 2016.

\bibitem{Jiang2017}
C.~Jiang, H.~Zhang, Y.~Ren, Z.~Han, K.-C. Chen, and L.~Hanzo, ``{Machine
  Learning Paradigms for Next-Generation Wireless Networks},'' \emph{IEEE
  Wireless Communications}, pp. 2--9, 2017.

\bibitem{Shafiq2011}
M.~Z. Shafiq, L.~Ji, A.~X. Liu, and J.~Wang, ``{Characterizing and modeling
  internet traffic dynamics of cellular devices},'' in \emph{Proceedings of the
  ACM SIGMETRICS joint international conference on Measurement and modeling of
  computer systems}.\hskip 1em plus 0.5em minus 0.4em\relax New York, New York,
  USA: ACM Press, 2011, p. 305.

\bibitem{Oliveira2016}
T.~P. Oliveira, J.~S. Barbar, and A.~S. Soares, ``{Computer network traffic
  prediction: a comparison between traditional and deep learning neural
  networks},'' \emph{International Journal of Big Data Intelligence}, vol.~3,
  no.~1, p.~28, 2016.

\bibitem{Cortez2012}
P.~Cortez, M.~Rio, M.~Rocha, and P.~Sousa, ``{Multi-scale Internet traffic
  forecasting using neural networks and time series methods},'' \emph{Expert
  Systems}, vol.~29, no.~2, pp. 143--155, dec 2012.

\bibitem{Huang2017}
C.-W. Huang, C.-T. Chiang, and Q.~Li, ``{A Study of Deep Learning Networks on
  Mobile Traffic Forecasting},'' in \emph{IEEE International Symposium on
  Personal, Indoor and Mobile Radio Communications (PIMRC)}, Montreal, Canada,
  Oct. 2017.

\bibitem{Saker2012}
L.~Saker, S.~Elayoubi, R.~Combes, and T.~Chahed, ``{Optimal Control of Wake Up
  Mechanisms of Femtocells in Heterogeneous Networks},'' \emph{IEEE Journal on
  Selected Areas in Communications}, vol.~30, no.~3, pp. 664--672, apr 2012.

\bibitem{chen2015energy}
X.~Chen, J.~Wu, Y.~Cai, H.~Zhang, and T.~Chen, ``Energy-efficiency oriented
  traffic offloading in wireless networks: A brief survey and a learning
  approach for heterogeneous cellular networks,'' \emph{IEEE Journal on
  Selected Areas in Communications}, vol.~33, no.~4, pp. 627--640, 2015.

\bibitem{Simsek2015}
M.~Simsek, M.~Bennis, and I.~Guvenc, ``{Context-aware mobility management in
  HetNets: A reinforcement learning approach},'' in \emph{2015 IEEE Wireless
  Communications and Networking Conference (WCNC)}.\hskip 1em plus 0.5em minus
  0.4em\relax IEEE, Mar. 2015, pp. 1536--1541.

\bibitem{Sun2015}
Y.~Sun, H.~Shao, X.~Liu, J.~Zhang, J.~Qiu, and Y.~Xu, ``{Traffic Offloading in
  Two-Tier Multi-Mode Small Cell Networks over Unlicensed Bands: A Hierarchical
  Learning Framewo},'' \emph{KSII Transactions on Internet and Information
  Systems}, vol.~9, no.~11, Nov. 2015.

\bibitem{Zhang2016}
S.~Zhang, N.~Zhang, S.~Zhou, J.~Gong, Z.~Niu, and X.~Shen, ``{Energy-Aware
  Traffic Offloading for Green Heterogeneous Networks},'' \emph{IEEE Journal on
  Selected Areas in Communications}, vol.~34, no.~5, pp. 1116--1129, May 2016.

\bibitem{Wei2018}
Y.~Wei, F.~R. Yu, M.~Song, and Z.~Han, ``{User Scheduling and Resource
  Allocation in HetNets With Hybrid Energy Supply: An Actor-Critic
  Reinforcement Learning Approach},'' \emph{IEEE Transactions on Wireless
  Communications}, vol.~17, no.~1, pp. 680--692, Jan. 2018.

\bibitem{mnih2015human}
V.~Mnih, K.~Kavukcuoglu, D.~Silver, A.~A. Rusu, J.~Veness, M.~G. Bellemare,
  A.~Graves, M.~Riedmiller, A.~K. Fidjeland, G.~Ostrovski \emph{et~al.},
  ``Human-level control through deep reinforcement learning,'' \emph{Nature},
  vol. 518, no. 7540, pp. 529--533, 2015.

\bibitem{Barlacchi2015}
G.~Barlacchi, M.~{De Nadai}, R.~Larcher, A.~Casella, C.~Chitic, G.~Torrisi,
  F.~Antonelli, A.~Vespignani, A.~Pentland, and B.~Lepri, ``{A multi-source
  dataset of urban life in the city of Milan and the Province of Trentino.}''
  \emph{Scientific data}, vol.~2, p. 150055, 2015.

\bibitem{sutton1998reinforcement}
R.~S. Sutton and A.~G. Barto, \emph{Reinforcement learning: An
  introduction}.\hskip 1em plus 0.5em minus 0.4em\relax MIT press Cambridge,
  1998, vol.~1, no.~1.

\bibitem{watkins1992q}
C.~J. Watkins and P.~Dayan, ``Q-learning,'' \emph{Machine learning}, vol.~8,
  no. 3-4, pp. 279--292, 1992.

\bibitem{marsan2012multiple}
M.~A. Marsan, L.~Chiaraviglio, D.~Ciullo, and M.~Meo, ``Multiple daily base
  station switch-offs in cellular networks,'' in \emph{Communications and
  Electronics (ICCE), 2012 Fourth International Conference on}.\hskip 1em plus
  0.5em minus 0.4em\relax IEEE, 2012, pp. 245--250.

\bibitem{soh2013energy}
Y.~S. Soh, T.~Q. Quek, M.~Kountouris, and H.~Shin, ``Energy efficient
  heterogeneous cellular networks,'' \emph{IEEE Journal on Selected Areas in
  Communications}, vol.~31, no.~5, pp. 840--850, 2013.

\bibitem{liu2016small}
C.~Liu, B.~Natarajan, and H.~Xia, ``Small cell base station sleep strategies
  for energy efficiency,'' \emph{IEEE Transactions on Vehicular Technology},
  vol.~65, no.~3, pp. 1652--1661, 2016.

\bibitem{siris2013performance}
V.~A. Siris and M.~Anagnostopoulou, ``Performance and energy efficiency of
  mobile data offloading with mobility prediction and prefetching,'' in
  \emph{World of Wireless, Mobile and Multimedia Networks (WoWMoM), 2013 IEEE
  14th International Symposium and Workshops on a}.\hskip 1em plus 0.5em minus
  0.4em\relax IEEE, 2013, pp. 1--6.

\bibitem{UlIslam2012}
M.~N. ul~Islam and A.~Mitschele-Thiel, ``{Cooperative Fuzzy Q-Learning for
  self-organized coverage and capacity optimization},'' in \emph{2012 IEEE 23rd
  International Symposium on Personal, Indoor and Mobile Radio Communications -
  (PIMRC)}.\hskip 1em plus 0.5em minus 0.4em\relax IEEE, Sep. 2012, pp.
  1406--1411.

\bibitem{Bennis2013}
M.~Bennis, S.~M. Perlaza, P.~Blasco, Z.~Han, and H.~V. Poor,
  ``{Self-Organization in Small Cell Networks: A Reinforcement Learning
  Approach},'' \emph{IEEE Transactions on Wireless Communications}, vol.~12,
  no.~7, pp. 3202--3212, Jul. 2013.

\bibitem{Chinchali2018}
S.~Chinchali, P.~Hu, T.~Chu, M.~Sharma, M.~Bansal, R.~Misra, M.~Pavone, and
  S.~Katti, ``{Cellular Network Traffic Scheduling With Deep Reinforcement
  Learning},'' in \emph{AAAI}, 2018.

\bibitem{Xu2016}
F.~Xu, Y.~Lin, J.~Huang, D.~Wu, H.~Shi, J.~Song, and Y.~Li, ``{Big Data Driven
  Mobile Traffic Understanding and Forecasting: A Time Series Approach},''
  \emph{IEEE Transactions on Services Computing}, vol.~9, no.~5, pp. 796--805,
  Sep. 2016.

\bibitem{Zhou2012}
X.~Zhou, Z.~Zhao, R.~Li, {Yifan Zhou}, and H.~Zhang, ``{The predictability of
  cellular networks traffic},'' in \emph{2012 International Symposium on
  Communications and Information Technologies (ISCIT)}.\hskip 1em plus 0.5em
  minus 0.4em\relax IEEE, oct 2012, pp. 973--978.

\bibitem{Li2017}
R.~Li, Z.~Zhao, J.~Zheng, C.~Mei, Y.~Cai, and H.~Zhang, ``{The Learning and
  Prediction of Application-Level Traffic Data in Cellular Networks},''
  \emph{IEEE Transactions on Wireless Communications}, vol.~16, no.~6, pp.
  3899--3912, jun 2017.

\bibitem{Lv2015}
Y.~Lv, Y.~Duan, W.~Kang, Z.~Li, and F.~Y. Wang, ``{Traffic Flow Prediction With
  Big Data: A Deep Learning Approach},'' \emph{IEEE Transactions on Intelligent
  Transportation Systems}, vol.~16, no.~2, pp. 865--873, 2015.

\bibitem{Huang2014}
W.~Huang, G.~Song, H.~Hong, and K.~Xie, ``{Deep architecture for traffic flow
  prediction: Deep belief networks with multitask learning},'' \emph{IEEE
  Transactions on Intelligent Transportation Systems}, vol.~15, no.~5, pp.
  2191--2201, 2014.

\bibitem{Ho2014}
C.~K. Ho, D.~Yuan, and S.~Sun, ``{Data Offloading in Load Coupled Networks: A
  Utility Maximization Framework},'' \emph{IEEE Transactions on Wireless
  Communications}, vol.~13, no.~4, pp. 1921--1931, apr 2014.

\bibitem{lin1993reinforcement}
L.-J. Lin, ``Reinforcement learning for robots using neural networks,''
  Carnegie-Mellon Univ Pittsburgh PA School of Computer Science, Tech. Rep.,
  1993.

\bibitem{OpenCellID}
\BIBentryALTinterwordspacing
\emph{OpenCellID}, 2018 (accessed March, 2018). [Online]. Available:
  \url{https://opencellid.org/}
\BIBentrySTDinterwordspacing

\end{thebibliography}

\begin{IEEEbiography}[{\includegraphics[width=1in,height=1.25in,clip,keepaspectratio]{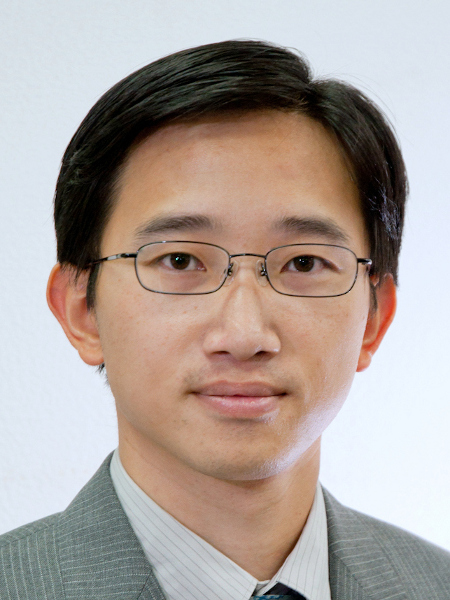}}]{Chih-Wei Huang}
received a B.S. degree from National Taiwan University, Taipei, in 2001, an M.S. degree from Columbia University, New York, in 2004, and the Ph.D. degree from University of Washington, Seattle, in 2009, all in electrical engineering.

He joined the Department of Communication Engineering, National Central University, Taoyuan, Taiwan, in 2010. He is currently an Associate Professor heading the Information Processing and Communications (IPC) laboratory. From 2006 to 2009, he was an intern researcher at Siemens Corporate Research and Microsoft Research. He is the author of papers in areas including wireless networking, multimedia communications, machine learning, digital signal processing, and information retrieval. Dr. Huang has received the best paper awards from IEEE ICC 2018 and WOCC 2015 conferences.
\end{IEEEbiography}

\begin{IEEEbiography}[{\includegraphics[width=1in,height=1.25in,clip,keepaspectratio]{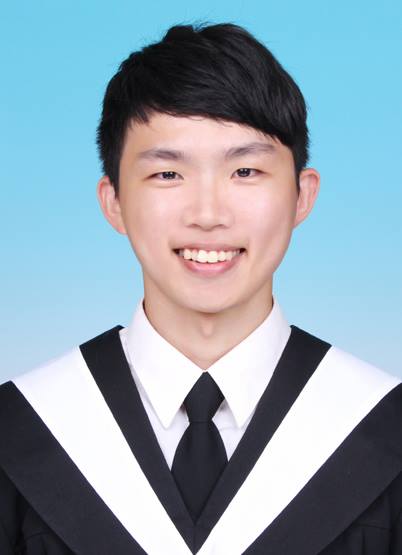}}]{Po-Chen Chen}
received a B.S. degree in Communication Engineering from the National Central University, Taoyuan, Taiwan, in 2017. He completed an M.S. degree at National Central University, in 2019 on radio resource management for 5G networks in Communication Engineering. He has been with the TSMC since 2019 and currently a R\&D engineer.
\end{IEEEbiography}

\end{document}